\documentclass[a4paper,fleqn]{cas-sc}

\usepackage[numbers]{natbib}

\def\tsc#1{\csdef{#1}{\textsc{\lowercase{#1}}\xspace}}
\tsc{WGM}
\tsc{QE}
\tsc{EP}
\tsc{PMS}
\tsc{BEC}
\tsc{DE}

\definecolor{red}{HTML}{F0684D}

\usepackage[super]{nth}
\usepackage{enumitem}
\newlist{operator}{enumerate}{1}
\setlist[operator]{label=${^\arabic*\Phi}$.}
\usepackage{float}
\usepackage{xurl}

\begin{document}
\let\WriteBookmarks\relax
\def\floatpagepagefraction{1}
\def\textpagefraction{.001}
\shorttitle{Exploring, browsing and interacting with multi-scale structures of knowledge}
\shortauthors{Quentin Lobbé, Alexandre Delanoë, David Chavalarias.}

\title [mode = title]{Exploring, browsing and interacting with multi-scale structures of knowledge}                      

\author{Quentin Lobbé}
\author{Alexandre Delanoë}
\author{David Chavalarias}

\address{CNRS, Complex Systems Institute of Paris Île-de-France (ISC-PIF), Paris, France}

\begin{abstract}
The ICT revolution has given birth to a world of digital traces. A wide number of knowledge-driven domains like science are daily fueled by unlimited flows of textual contents. In order to navigate across these growing constellations of words, interdisciplinary innovations are emerging at the crossroad between social and computational sciences. In particular, complex systems approaches make it now possible to reconstruct multi-level and multi-scale structures of knowledge by means of \textit{phylomemies}: inheritance networks of elements of knowledge. 

In this article, we will introduce an endogenous way to visualize the outcomes of the phylomemy reconstruction process by combining both synchronic and diachronic approaches. Our aim is to translate high-dimensional phylomemetic networks into graphical projections and interactive visualizations. To that end, we will use seabed and kinship views to translate the multi-level and multi-scale properties of complex branches of knowledge. We will then define a generic macro-to-micro methodology of exploration implemented within an open source software called \textit{Memiescape} and validate our approach by browsing through the reconstructed histories of thousands of scientific publications and clinical trials.   
\end{abstract}




\maketitle

%
%

\section{Introduction}
\label{introduction}

Since the dawn of humanity, writing has been one of the first mnemotechnology: a technique not only designed to fix a thought on a medium but also a dynamic tool for the elaboration of a collective memory \citep{stiegler1998leroi}. Written texts can thus be considered as vectors of knowledge as well as providers of socio-historical contexts. The accumulation of Mesopotamian clay tablets (4000~BC) or the elaboration of the Vivarium library (535~AD-555~AD) gave early evidence of a growing will to collect and provide access to isolated elements of knowledge. Later on, with the transition from manuscript to book \citep{febvre2013apparition}, textual contents outgrew erudite communities and started to touch all layers of the population, up to the present day: we are now daily fueled by unlimited flows of articles, novels, messages, tweets, etc. The recent ICT revolution~\citep{borgman2003gutenberg} has given birth to an unprecedented world of digital traces and has impacted a wide number of knowledge-driven domains such as education or policy making.

Science, in particular, has been one of the first area to experiment this digital shift. Databases of scientific publications are scaling up and it is now possible to dive into the amazing richness of most of these catalogs. Qualitative sciences are also taking advantage of the ICT revolution by integrating large cultural data sets (digitized historical documents, social networks footprints, archived Web sites, etc.) within their own scopes of analysis \citep{lobbe2018dead}. Digital-born fields of research have thus emerged at the crossroad between social and computational sciences. But whether we speak of digital humanities \citep{rogers2013digital} or cultural analytics \citep{manovich2015science}, it seems that all these domains end up facing the same issue: how to navigate across growing constellations of words and texts?

As early as the \nth{18} century, while he was completing the first edition of the \textit{Encyclopédie} \citep{d1894discours}, d'Alembert suggested the idea of using trees to situate the future reader: 

\begin{quotation}
\textit{``[...] above this vast labyrinth, whence he can perceive the principal sciences and the arts simultaneously. From there he can see at a glance the objects of their speculations and the operations which can be made on these objects; he can discern the general branches of human knowledge, the points that separate or unite them; and sometimes he can even glimpse the secrets that relate them to one another [...]''}\footnote{d'Alembert, Jean-Baptiste le Rond. "Preliminary Discourse." The Encyclopedia of Diderot \& d'Alembert Collaborative Translation Project. Translated by Richard N. Schwab and . Ann Arbor: Michigan Publishing, University of Michigan Library, 2009. Web. 2021. \url{http://hdl.handle.net/2027/spo.did2222.0001.083}. Trans. of "Discours Préliminaire," Encyclopédie ou Dictionnaire raisonné des sciences, des arts et des métiers, vol. 1. Paris, 1751.}
\end{quotation}

Likewise, at the beginning of the 1900s, Paul Otlet planned to use experimental interfaces for the legendary \textit{Mundaneum} \citep{manfroid2010paul}. The visitors of the library were supposed to have access to the collected documents’ references by means of macro-visualizations\footnote{\url{https://upload.wikimedia.org/wikipedia/commons/2/29/Mondoth\%C3\%A8que_02.jpg}} placed on top of mobile pieces of furniture. These few examples are evidence of a historical need: researchers have always been looking for tools able to capture and visualize overall representations of wide elements of knowledge. In line with D’Alembert’s vision, our article’s purpose is to give researchers the means to interact with large corpora of texts in relevance with their investigations. To that end, we will rely here on the most recent developments in the fields of \textit{collective intelligence} and \textit{reconstruction methods}.

\paragraph{Collective shapes of knowledge.} Because there are structures inside knowledge, a given text can always be studied in relation to others or in light of a specific socio-cultural context. By way of textual traces, human beings are calling out to one another: citations, retweets, controversy, etc. We are the architects of a giant web of elements of knowledge whose very structures and shapes convey information of their own \citep{chavalarias2019formes}. Like ants or bees, through the aggregation of individual contributions, we collectively achieve complex tasks that are out of the reach of individuals. This phenomenon is called \textit{collective intelligence} \citep{bonabeau_intelligence_1994} and relies on a core mechanism called \textit{stigmergy}; that is, the indirect coordination between an agent and an action through the environment \citep{theraulaz1999brief}. From scientific archives to Web pages and online ratings, our digital societies are literally embedded in a stigmergic environment. Wikipedia, for instance, is an emblematic example \citep{lee2017}. The global shapes of these traces of collective intelligence constitute a full-fledged source of knowledge.

\paragraph{Reconstruction methods.} Nowadays, complex systems approaches enable us to reconstruct the collective shapes and ontogeny of large corpora of texts. We call \textit{reconstruction methods} all techniques implemented to understand a complex object or natural phenomenon by means of both the observation of patterns and the analysis of processes. Such methods are part of the larger family of \textit{phenomenological reconstructions}, designed to find reasonable approximations of the structure and dynamics of a given phenomenon \citep{bourgine_formal_2009}. Reconstruction methods can be summarized by the generic workflow $O \rightarrow R \rightarrow V$, where $O$ represents a complex object associated to a set of properties. Based on a collected data set,$O$ is next reconstructed as a formal object $R$ described in a high-dimensional space on the basis of a collected data set. The process ends with the dimensional reduction of $R$, so that it can be projected as a human-readable visualization~$V$. In this paper, we focus on the \textit{phylomemy reconstruction} method (see \ref{phylomemy_reconstruction}) which consists in reconstructing inheritance networks of elements of knowledge on top of timestamped corpora of textual documents~\citep{chavalarias2013phylomemetic}. But while the most recent research have investigated the multi-level and multi-scale properties of phylomemies in~$R$~\citep{chavalarias2020}, the question of their visualization in~$V$ remains an open challenge. How can we explore, browse and interact with wide elements of knowledge through a phylomemy? How can we translate their inner processes in a graphical way? 

\paragraph{Summary of main contributions.} We'll first review the state of the art of \textit{text analysis and knowledge visualization} methods (see \ref{knowledge_visualization}). We will then focus on the field of \textit{science dynamics} (see \ref{science_dynamics}) before drilling down through the phylomemy reconstruction method (see \ref{phylomemy_reconstruction}). In doing so, we will highlight the inner properties and components of the phylomemies we aim to visualize. After having investigated the notions of synchrony and diachrony, we will define two complementary axes of visualization: the seabed view and the kinship view (see \ref{views}), and discuss ways to interact with them through micro, mezzo and macro lenses of exploration (see \ref{lenses}). Using the free software \textit{Gargantext}, we will then develop our approach thanks to a new open source vizualisation system called \textit{Memiescape} (see~\ref{implementation}) and validate it by browsing through the historical landscapes of various domains of research (see \ref{results}). We will end up by introducing a generic methodology for browsing through multi-scale structures of knowledge (see \ref{discussion}).

%
%

\section{State of the art and insights}

Mapping large corpora of texts is an interdisciplinary domain of research that has recently expended under the influence of the data revolution \cite{kitchin_data_2014}. In order to position our contribution, we will adopt a \textit{co-word analysis} approach~\citep{callon1991co, delanoe2014dematerialization, delanoe_mining_2020} to map out the scientific literature of \textit{text analysis and knowledge visualization}. To that end, let us start by framing the domain with a complex query (see \ref{appendix_literature}) based on generic terms such as \textit{science map}, \textit{information cartography}, \textit{knowledge map}, etc. With it we extract a corpus of meta-data\footnote{The corpus and list of terms can be downloaded in \cite{DVN/SLARHQ_2021}} from 13844 documents (titles and abstracts published between the~'80s and the present days) from the \textit{Web of Science} (Wos) online database and called $\mathcal{D}_{maps}$. 

\begin{figure}[!h]
 \centering
 \includegraphics[width=\linewidth]{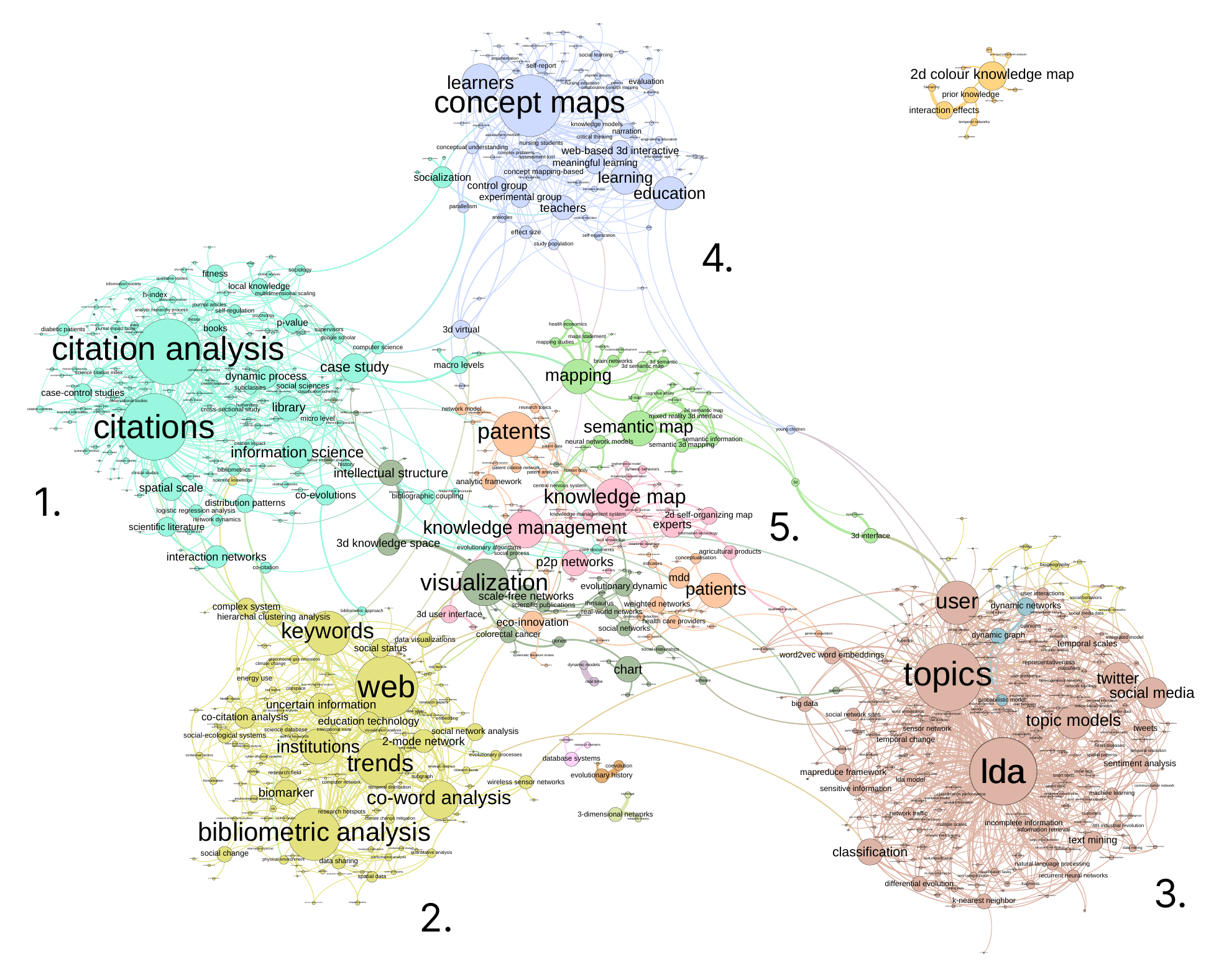}
 \caption{Map of the meta-data of 13844 publications related to \textit{text analysis and knowledge visualization}, extracted from the Web of Science by using a confidence similarity metric. Generated with Gargantext and spatialized with Gephi. Interactive version available at \url{http://maps.gargantext.org/maps/sciencemaps/}.}
 \label{fig:map_knowledge_visualization}
\end{figure}

Based on both automatic extraction and human pruning, the free software Gargantext (see \ref{gargantext}) identifies a list of 876 terms constituting the specific vocabulary enclosed in our targeted corpus. The software then generates a semantic landscape appearing as a network of terms (i.e., nodes). In addition, the weighted edges represent any similarities found between terms; they are processed regarding the terms' co-occurrences in the corpus. We have elected the \textit{confidence} similarity metric \citep{dias_mapping_2008}, a good proxy to measure direct interactions between elements of vocabulary. A map based on the confidence metric might indeed reveal the topology of the relationships between \textit{hypernymies} (i.e., superordinate grouping terms) and \textit{hyponymies} (i.e., terms more specific than others), both in terms of meanings and uses. Such a map is thus able to reconstruct a semantic area made of general concepts linked to more specific notions. In our case, a community detection algorithm also highlights the main research domains \citep{blondel_fast_2008} as displayed by \autoref{fig:map_knowledge_visualization}.

\subsection{The scientific landscape of text analysis and knowledge visualization}
\label{knowledge_visualization}

Let us now examine map \ref{fig:map_knowledge_visualization} where five main communities of research (the numbered dense areas of terms) appear. We here make the assumption that a scientific community can be represented by the use of specific and shared elements of vocabulary. We review these communities by considering their most representative publications:

\newpage

\begin{enumerate}

  	\item \textbf{Citation analysis.} It was the core of \textit{scientometrics} in the 1970s \citep{garfield1972citation} and was used as a method to assess \textit{scientific impact}. Although focusing on \textit{knowledge domains}, it never really dealt with vizualisation and left it to the \textit{bibliometrics} community.
 
	\item \textbf{Bibliometrics analysis.} In addition to pure citation analysis, the field of \textit{bibliometrics analysis} developed in the early 1970s, in keeping with \textit{bibliographic coupling} and \textit{co-citation} techniques \citep{kessler_bibliographic_1963,small1973co}. Later on, following the creation of the \textit{Web}, bibliometrics approaches enjoyed a surge of interest with the emergence of \textit{hyperlinked data}~\citep{kleinberg_authoritative_1999}. \textit{Visualization analysis} became central since it offered tools to describe \textit{conceptual structures} of science such as \textit{research fronts}, \textit{hot topics} and \textit{trends}, etc. \citep{white_visualizing_1998,borner_visualizing_2003,chen_citespace_2006} -- which might also be studied together as socio-semantic networks \citep{roth_social_2010}. 
	
	\textit{Co-word analysis} is a bottom-up approach. First developed by sociologists in the 1980s \citep{callon1986mapping} to reconstruct the dynamics of \textit{research themes} out of words' \textit{co-occurrence}, it quickly paved the way to hybrid research \citep{braam_mapping_1991,boyack_co-citation_2010}.
	
	All these methods have primarily borrowed concepts from \textit{graphs} and \textit{social networks} analysis. The sub-field of \textit{science mapping} aims to explore the \textit{social structures} and \textit{temporal evolution} of \textit{academic research}~\citep{small1999visualizing} with the help of computer science techniques \citep{harries2004hyperlinks}. 
	
	Nowadays, \textit{science maps} are interdisciplinary objects of research resulting from both quali-quantitative and socio-technical processes \citep{chen2019visualizing}. The growth of scientific databases has finally stimulated the visualization of wide \textit{citation landscapes} \citep{small1997update} or complex atlases of sciences~\citep{borner2010atlas}.
	
	
	\item \textbf{Topic modeling.} This field emerged in the early 2000s at the instigation of a community of statisticians who first used the \textit{Latent Dirichlet Allocation} method \citep{blei2003latent} to characterize \textit{collections of documents}. Although focusing primarily on \textit{document classification} \citep{wei_lda-based_2006}, \textit{recommendation} \citep{wang_collaborative_2011} or \textit{sentiment analysis} \citep{lin2009joint}, parts of their most recent works have started to investigate science mapping \citep{millar2009document, yang2017vistopic}.
	
	\item \textbf{Concept and semantic maps.} In the 1990s, both mapping techniques gained entry to the broader field of \textit{science of education}  \citep{novak1990concept} as means to support \textit{knowledge integration} \citep{jonassen1997concept}. Concept mapping has been deeply influenced by psychology and cognition. A \textit{concept map} can be defined as a \textit{graphical representation} designed to highlight the relationships between ideas or \textit{key concepts} \citep{kinchin2000concept}. Its purpose is to clarify a given topic as well as its underlying \textit{cognitive structure} \citep{rohrer1998,benevene2017representation} by means of \textit{ontologies}, \textit{mind maps}, \textit{mental models}, etc. Unlike co-word approaches, \textit{concept maps} were initially supposed to translate elements of knowledge issued by \textit{learners} and \textit{teachers} in a top-down way. But the recent influence of data mining methods has reversed this trend by increasing the use of bottom-up recommendation systems or topic detection, along with the introduction of \textit{visualization tools}~\citep{nesbit2006learning, romero2007educational,slater2017tools}.	
	
	\item \textbf{Domains with peripheral concerns.} Unlike clusters no.1 to no.3, the communities represented by cluster no.5 are not focusing on a single method, but rather borrow existing techniques from the latter to study their own objects of research. Among these peripheral domains, the fields of \textit{knowledge management} \citep{lin2006knowledge}, \textit{business intelligence} \cite{pyo2005knowledge} and \textit{patent analysis} \citep{tseng2007text} stand out. 
	
	Yet the scope of cluster no.5 remains unclear. It gathers methodologies popularized by mathematicians and computer scientists (\textit{community detection}, \textit{scale free networks}, \textit{small world networks}, etc.) as well as wide and transversal scientific domains  (\textit{visualization}, \textit{mapping}, \textit{data science}, etc.) In fact, as text analysis and knowledge visualization are not really at the heart of communities no.1 to no.4’s concerns, our initial query (see \ref{appendix_literature}) only captured a slice of their related literature, hence the sparse and fragmented aspect of the cluster no.5 in comparison.
	 
\end{enumerate}

\subsection{Revealing the dynamics of science}
\label{science_dynamics}

For our part, we choose co-word approaches (\autoref{fig:map_knowledge_visualization}, cluster no.2) underlied by phylomemy reconstruction (see \ref{phylomemy_reconstruction}). Let us now focus on how phylomemies deal with and effectively translate the temporal aspect of the evolution of knowledge.

Over the last decade, science has become a fertile field of investigation for the study of time-related dynamics. Being able to access countless digitized academic archives has acted as an incentive for researchers \citep{chen_towards_2009, zeng_science_2017}. Each of the communities appearing on \autoref{fig:map_knowledge_visualization} has developed its own set of temporally-aware techniques, from citations dynamics~\citep{ramos-rodriguez_changes_2004, chen_citespace_2006, chen_towards_2009, rosvall_mapping_2010} to topic modeling over time \citep{wang_topics_2006, wang_continuous_2015, yang2017vistopic, cui_how_2014, minjeong_topiclens:_2017} and co-word evolution in the case of phylomemies \citep{chavalarias2013phylomemetic, rule_lexical_2015}. The latest developments on phylomemy reconstruction \citep{chavalarias2020} have indeed enriched these seminal approaches: by means of innovative multi-level properties articulated with an intuitive parameter of resolution (see \ref{phylomemy_reconstruction}), the resulting high dimensional object in $R$ encompasses several scales of aggregation, paving the way to a new generation of visualizations in $V$.

But regardless of the field of application, taking both the multi-level structure and the inner dynamics of scientific domains into account in an interactive visualization system remains an open challenge; and none of the aforementioned research has actually succeeded in representing the two aspects at the same time in an effective way. For example: while stream graphs successfully convey the evolution of topics, they fail to translate at the same time the structural relationships between these topics \citep{ rule_lexical_2015, yang2017vistopic}. Alluvial diagrams and variants likewise hit a brick wall when it comes to revealing the connection between branches of knowledge when these are too complex to be interpreted clearly~\citep{rosvall_mapping_2010, chavalarias2013phylomemetic, cui_how_2014, shahaf_information_2013}. 

\paragraph{Underway processes.} It is our assumption that multi-views approaches \citep{cuenca2018, kim2016topiclens, zhang2016} might be the missing key to visualizing high-dimensional objects. But the originality and richness of a phylomemy might also be impoverished by the use of classical representations. The reconstruction process of a given phylomemy phylomemy in and of itself indeed provides information as valuable as the very dynamics and structures it aims to reveal. With this in mind, we can therefore draw inspiration from any research that focuses on the visualization of processes under way. For instance, recent breakthroughs in terms of writing processes have enabled researchers to design ad hoc representations which endogenously translate the making of a text as well as its inner content~\citep{Perez2018}. Similarly, in what follows, we plan to build a pair of dedicated views organically deriving from the phylomemy reconstruction process and, by doing so, to translate the phylomemy's multi-level and multi-scale properties.

\subsubsection{The phylomemy reconstruction process}
\label{phylomemy_reconstruction}

The phylomemy reconstruction process \citep{chavalarias2013phylomemetic} is part of the larger family of reconstruction methods (see \ref{introduction}) and is based on co-word analysis approach (see \ref{knowledge_visualization}). Within the scope of the generic chain $O \rightarrow R \rightarrow V$, a phylomemy is a formal object $R$ designed to reveal conceptual lineages out of any kind of unstructured but timestamped corpora of texts $O$. Here, the transition $\Phi$ from $O$ to $R$ can be described as a combination of four successive operators of reconstruction $\Phi={^4\Phi} \circ {^3\Phi} \circ {^2\Phi} \circ {^1\Phi}$ \citep{chavalarias2020}, where:

\begin{figure}[!h]
 \centering
 \includegraphics[width=\linewidth]{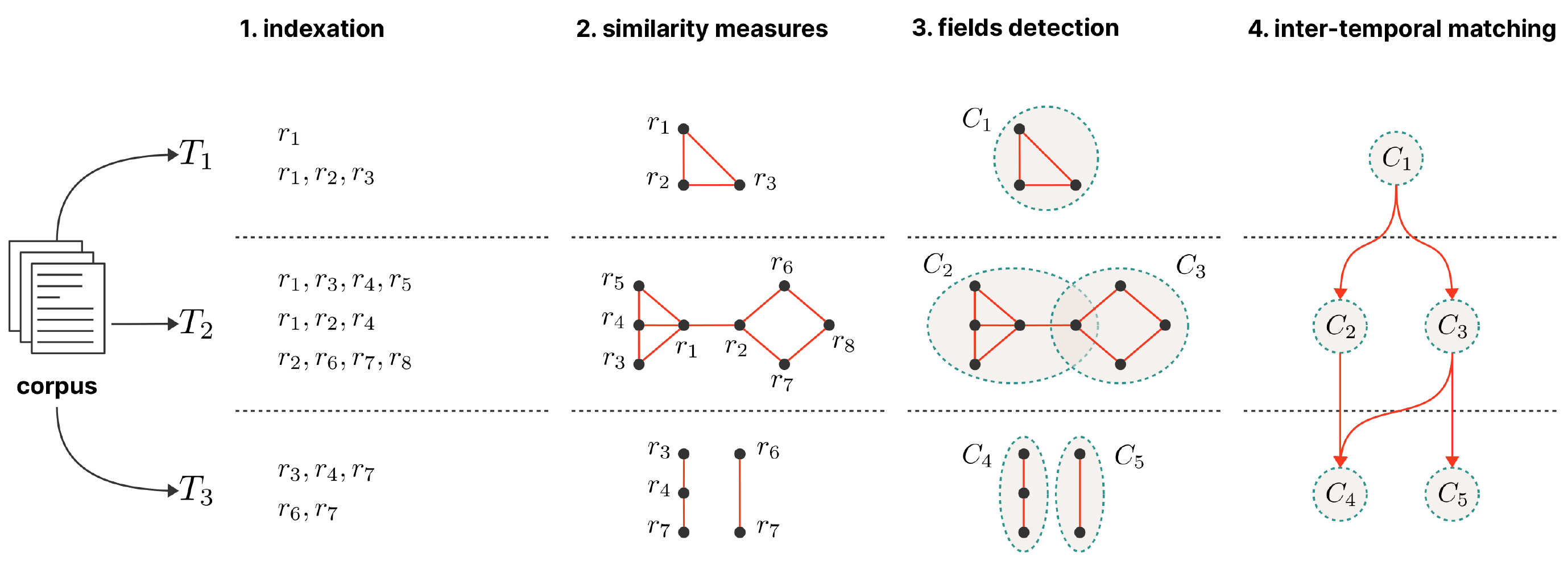}
 \caption{The four operators of reconstruction}
 \label{fig:operators}
\end{figure}

\begin{operator}

	\item \textbf{Indexation.} We start by framing a corpus of textual documents from a given database (see \ref{results}) before extracting its core vocabulary as a list $\mathcal{L}=\{r_i\ | i \in \mathcal{I}\}$ of root terms $r_i$. We then choose a temporal resolution (e.g., 3 years) that discretizes our corpus into an ordered set of periods $\mathcal{T^*} = \{T_i\}_{1 \leq i \leq K}$,  $T_i\subset \mathcal{T}$. For each period, we finally compute the co-occurrence of terms in the documents.

	\item \textbf{Similarity measures.} We build upon the matrix of co-occurrences to compute a similarity measure (e.g., the confidence measure, see \ref{knowledge_visualization}) that we use as a proxy to estimate the semantic similarity between the terms. 

	\item \textbf{Fields detection and clustering.} Within each period, the completion of ${^2\Phi} \circ {^1\Phi}$ results in a graph of similarity potentially containing meaningful sub-units of terms ${C}^T$ called \textit{fields}. We then use of specific clustering algorithms (like \textit{frequent item set} or \textit{maximal clique} \citep{chavalarias2020}) to identify these fields as sets of groups of terms over the periods, such as $\mathcal{C^*}=\{\mathcal{C}^T|T \in \mathcal{T^*}\}$ with $\mathcal{C}^T=\{C_j|j\in J^{T} \}$ and $C_j=\{r_i\ | r_i\in \mathcal{L},  i \in \mathcal{I}_j\ \subset \mathcal{I}\}$.

	\item \textbf{Inter-temporal matching.} Finally, an inter-temporal matching mechanism reconstructs any kinship connections between groups from one period of time to another. It tries to assign each group a set of parents and children and, by doing so, highlights elements of conceptual and semantic continuity over time called \textit{branches}. This resulting structure of terms, groups, links and branches determines the overall shape of the phylomemy.  

\end{operator}

\autoref{fig:operators} summarizes linkages between ${^1\Phi}, {^2\Phi}, {^3\Phi}$ and ${^4\Phi}$. But the comprehension of the last mechanism requires a fine-grained explanation. Indeed, ${^4\Phi}$ first relies on an inter-temporal matching function that derives from a similarity measure $\Delta: \mathcal{C} \times \ \mathcal{P}(\mathcal{C}) \rightarrow [0,1]$ (such as a \textit{Jaccard coefficient} \citep{dias_mapping_2008}). This function aims to create upstream/downstream kinship connections between a given group $C^T$ at period $T$ and any single/pairs of candidates~$\{C_j\}_j\subset \mathcal{P}(\mathcal{C})$ belonging to a strictly anterior/superior period $T'$. The resulting lineages are then validated if and only if their corresponding values of $\Delta$ satisfy a fixed threshold $\delta \geq 0$.

\begin{figure}[!h]
 \centering
 \includegraphics[width=\linewidth]{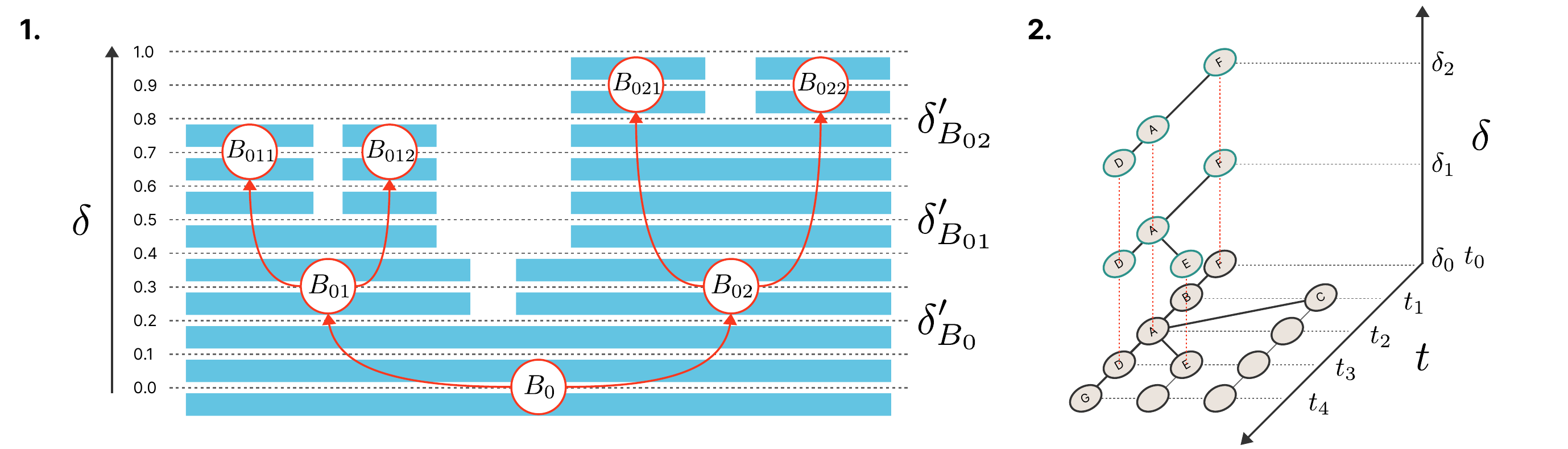}
 \caption{1. The sea level rise algorithm (the initial branch $B_0$ breaks into smaller branches $B_{011},B_{012},B_{021},B_{022}$) and 2. the resulting foliation (a temporal serie of clusters $[A...G]$ is associated to each value of $\delta$)}
 \label{fig:sea_level_rise}
\end{figure}

\paragraph{Sea level rise.} Threshold $\delta$ follows the completion of a \textit{sea level rise} algorithm \citep{chavalarias2020} which makes it gradually evolve, as explained by \autoref{fig:sea_level_rise}.1. At a low level of similarity $\delta = 0$, all groups are connected as part of a wide continent $\varphi_0$. Later on, the gradual rise of $\delta$ splits this land into smaller islands $\varphi_\delta$ by cutting some of the weakest inter-temporal links. Some branches become siblings while others split up earlier and then evolve separately. The newly-formed branches eventually drift away from each other, giving shape to the whole phylomemy: in other words a \textit{foliation on a temporal series of clustering} (see \autoref{fig:sea_level_rise}.2). In addition, the elevation of $\delta$ is locally and recursively scheduled within each of the resulting branches by a global function of quality $F_\lambda$. This function is based on notions pertaining to information retrieval called \textit{accuracy} and \textit{recall}\footnote{\textit{Accuracy} is the proportion of relevant elements among all the elements retrieved and \textit{recall} is the proportion of relevant elements actually retrieved among all the relevant ones.}. $F_\lambda$ determines whether a given branch should continue to be divided or not at the current value of $\delta$. To that end, the quality score is set up by a parameter $\lambda$ to predetermine the desired shape of the phylomemy: a continent (i.e., one large branch) or an archipelago of elements of knowledge (i.e., many small branches). The estimation of $\lambda$ is left to the researcher's discretion in light of her own expertise and research questions, which makes any phylomemy an artifact of the researcher's perception. 

\paragraph{Level and scale.} The phylomemy reconstruction process deals with the notions\footnote{For a full-length article on notions of level and scale, please look at \cite{chavalarias2020}.} of both \textit{level} and \textit{scale}. In complex systems approaches, levels are generally higher descriptors than scales. Here, the choice of $\lambda$ first establishes a \textit{level of observation}, namely the range of intrinsic complexity determining the structural and dynamical properties of the phylomemy. Within each level of observation, a \textit{phylomemetic network} $\varphi_\delta$ following the last and local elevations of $\delta$ can be extracted (e.g, the set of branches $B_{011},B_{012},B_{021},B_{022}$ in the figure \autoref{fig:sea_level_rise}.1). This foliation slice has its own \textit{scales of description}, i.e the range of resolution adopted to describe the structural organization of $\varphi_\delta$ (e.g, the various relations between terms, groups, branches, etc).

%
%

\section{Materials and methods}
\label{methods}

We will now focus on the last step of the reconstruction process $O \rightarrow R \rightarrow V$: our visualization system. Once we have reconstructed a phylomemy for a specific level of $\lambda$, we must reduce the dimensional complexity of the phylomemetic network $\varphi_\delta$ before projecting its \textit{multi-scale} properties onto a graphical space. To that end, we articulate four operators of projection: framing the \textit{graphical scope} (see~\ref{scope}), choosing the corresponding \textit{axis of navigation} (see~\ref{axis}), designing their related \textit{views} (see~\ref{views}) and building the \textit{lenses} of exploration (see \ref{lenses}). 

To illustrate our methodology, we’ll use a set of screenshots taken from \textit{Memiescape} (a working demonstrator, see~\ref{implementation}) showing the phylomemy of a heterogeneous corpus of meta-data in scientific publications called $\mathcal{D}_{cnrs}$ (see \autoref{fig:memiescape}). This corpus gathers 6000 top cited papers extracted from the \textit{Web of Science} (WoS). Written between the 1980s and the present day, at least one of each publication’s authors is affiliated to the french CNRS\footnote{Corpus and list of terms have been defined in \cite{chavalarias2020} and can be downloaded in \cite{DVN/SBH3EI_2021}}.

\subsection{Graphical scope}
\label{scope}

In order to translate our object of study into graphical terms, we first need to review the various elements we plan to visualize. Once we have identified the interconnections between these elements as well as any mutual constraints, we can build an actually endogenous representation.

The main elements of a phylomemy are \textit{terms}, \textit{groups} and \textit{branches}. They are subject to the structural constraint: $\mbox{terms} \in \mbox{groups} \in \mbox{branches}$. In addition, terms and branches evolve through time, from one period to another (forward and backward), by means of \textit{kinship} lines that connect pairs of groups together. These weighted connections result from the inter-temporal matching mechanism and thus convey a similarity score. 

But some of the kinship lines might have been cut off by the \textit{sea level rise} algorithm (see \ref{phylomemy_reconstruction}). We call these cut-off lines \textit{ghost lines} of the branches' drift. These artifacts are vectors of information: they convey the similarity gap between two consecutive branches and more specifically between their respective semantic contents. In addition, we are able to access the local range of elevation of each branch to deduce their highest level of $\delta$. By using ghost lines and elevation ranges, we can reconstruct the whole drifting history of the phylomemy's branches as a naturally hierarchical process.

Unlike terms, which might appear over and over again, groups are strictly timestamped within specific branches. But their corresponding dates give us the possibility to enrich each term with dynamical properties that spread along kinship lines or across distant branches. By doing so, we can determine the terms’ frequency of appearance per period or ask whether one of them is emerging or decreasing:

\begin{itemize}

\item \textbf{Emerging.} A term is emerging  if it appears at a specific period for the first time in the whole phylomemy.

\item \textbf{Decreasing.} A term is decreasing if it is used at a specific period for the last time in the whole phylomemy.

\end{itemize} 

\subsection{Axis of navigation}
\label{axis}

We now have to make sense of all these elements from a graphical perspective, bearing in mind that fitting too many dimensions into a single representation can decrease the quality of information displayed \cite{bertin1973traitement}. We therefore propose to  introduce complementary axes of navigation (foreshadowed in \ref{science_dynamics}) by using Saussure's linguistic concepts of \textit{synchrony} and \textit{diachrony} \cite{de1916cours}:

\begin{itemize}

\item \textbf{Synchrony.} Analyzing a language at a particular moment of its history.

\item \textbf{Diachrony.} Analyzing the historical and temporal evolution of a language.

\end{itemize}

Applied to the exploration of a phylomemy, a synchronic axis of navigation might highlight the timeless relations of similarity between branches while a diachronic axis may offer insight on the temporal evolution of the vocabulary. The above-mentioned elements can then be distributed along both axes as summarized by~\autoref{tab:synchrony_diachrony}. 

\begin{table}[width=.9\linewidth,cols=3,pos=h]
\caption{Distribution of the elements, connections and properties by axis of navigation}\label{tab:synchrony_diachrony}
\begin{tabular*}{\tblwidth}{@{} LLL@{} }
\toprule
& Synchrony & Diachrony\\
\midrule
elements & branches & terms, groups, branches \\
connections & ghost lines & kinship lines \\
properties & branches elevation, branches drift & groups dates, terms dynamics, terms frequency \\
\bottomrule
\end{tabular*}
\end{table}  

Based on this table, we can now translate all the phylomemetic elements graphically. The foundations of our graphical language (i.e. symbols, colors, sizes, etc.) has been inspired by the general principles of \textit{graphic} stated by Bertin~\citep{bertin1973traitement}, but its global organization and interactive mechanisms comes from our own understanding of the phylomemy reconstruction process.

\subsection{Complementary views}
\label{views}

We call \textit{views} a set of dedicated visualizations designed to address different issues but which can still be articulated together as part of a whole phylomemy. We here use two complementary views that respectively build upon the synchronic and diachronic axes of navigation. \autoref{fig:memiescape}.1 shows how we display these views on top of the other.

\subsubsection{The seabed view}

\begin{figure}[!h]
 \centering
 \includegraphics[width=\linewidth]{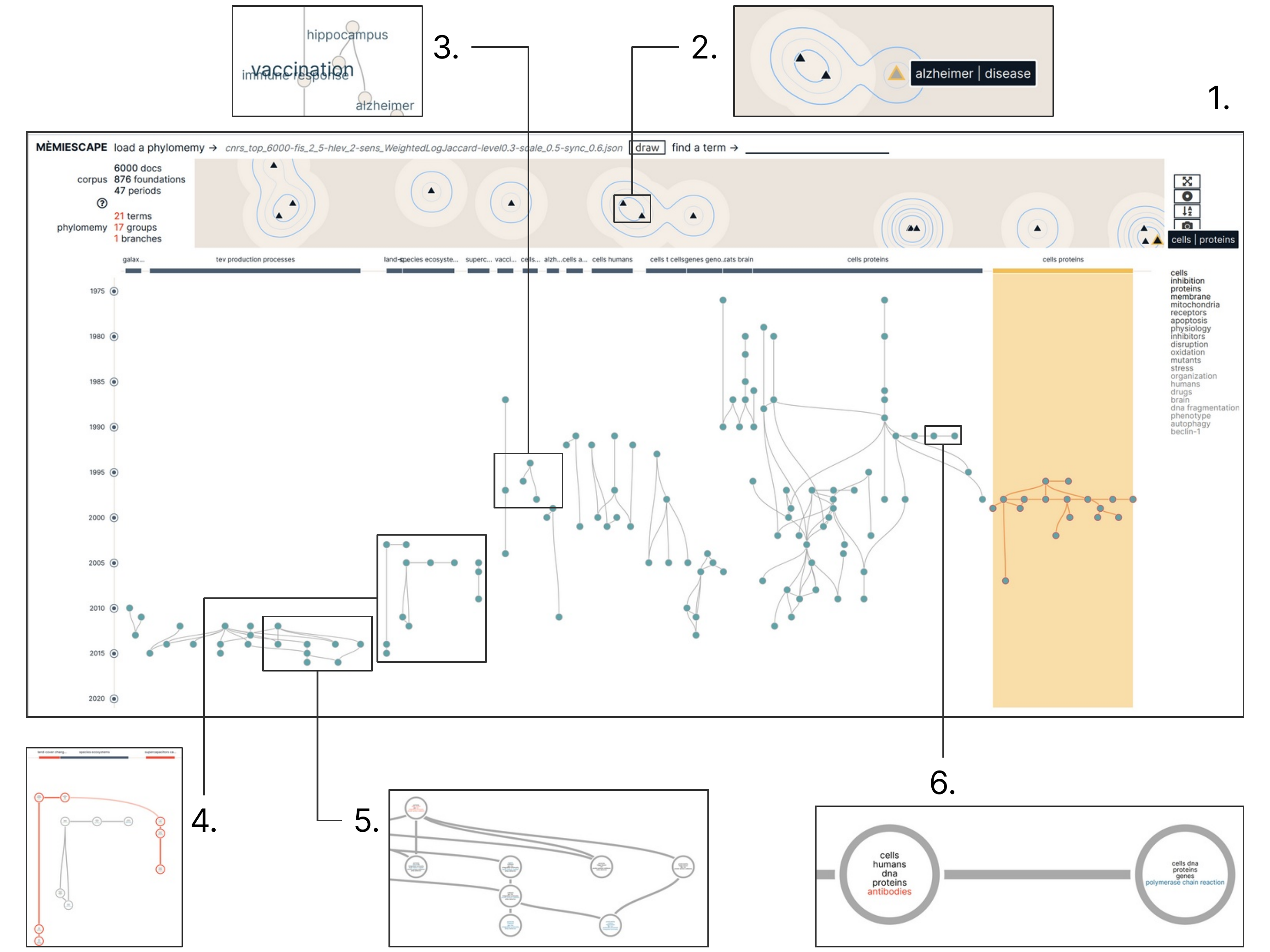}
 \caption{Screenshots taken from \textit{Memiescape} to visualize the phylomemy of the corpus $\mathcal{D}_{cnrs}$}
 \label{fig:memiescape}
\end{figure}

The first view (\autoref{fig:memiescape}.1, top part) follows a synchronic axis of navigation and aims to visualize the relations of similarity between branches\footnote{Note that regarding the whole reconstruction process (see \ref{phylomemy_reconstruction}) branches are temporally dense elements of knowledge with relations of similarity inherited from dynamical mechanisms, but here we choose to represent them without any explicit notion of time.}. We call this view the \textit{seabed view} as it extends the metaphor of the \textit{sea level rise} algorithm. The goal here is to project the branches in a two-dimensional topographic space. 

We only display the upper parts of the branches and use black triangles to symbolize their \textit{peaks} (\autoref{fig:memiescape}.2). In our system of coordinates, the ordinate goes from $0$ to $1$ (from top to bottom) and maps the last $\delta$ elevation of each branch. The abscissa translates the smallest gap of similarity between two consecutive branches. The seabed view thus helps us to understand a posteriori the outcomes of the sea level rise algorithm and its hierarchical drift: two branches displayed on opposite sides are certain to share no terms. In the \autoref{fig:memiescape} for instance, branches related to space exploration, dark matter and galaxies lie on the far left, while branches linked to genome, DNA fragmentation and human cells stay on the far right. In the same way, a branch that ends up at the bottom of the seabed view can be identified as a very specialized branch of knowledge (i.e., it results from a high level of $\delta$).

Finally, we draw a set of isolines around the branches' peaks (\autoref{fig:memiescape}.2, blue curves) by using their new spacial density\footnote{We use the library \url{https://github.com/d3/d3-contour} that relies on a marching squares algorithm \cite{maple2003geometric}}. The more branches we find side by side, the more isolines we draw. This method is an endogenous way to quickly highlight archipelagos of closely related branches like those focusing on \textit{alzheimer} and \textit{hippocampus} in the \autoref{fig:memiescape}.2. It gives us a hint about the way the same phylomemy could be shaped under different levels of observation by foreshadowing what mergers could occur between branches for lower values of $\lambda$.

\subsubsection{The kinship view}

In the \textit{Origin of Species} \citep{darwin2009origin}, Darwin used a single illustration\footnote{\url{https://commons.wikimedia.org/wiki/File:Darwin_divergence.jpg}} to suggest that evolution extends along lines throughout time. What he drew was \textit{evolutionary tree} where kinship connections were supposed to leave genetic information from ancestors to future generations. We now propose to use similar trees to represent the full body of our phylomemy's branches and explore the dynamics of knowledge at work. Tree-like visualizations already have a long-established legitimacy, from biology to anthropology~\citep{rivers1910genealogical} as vectors of comparative studies and topographic analysis. Our second view (\autoref{fig:memiescape}.1, bottom part) is therefore called the \textit{kinship view} and follows the diachronic axis of navigation.

We have chosen to use the same abscissa coordinates as in the seabed view to sort branches from left to right. We try to maintain their horizontal drifting gap as much as possible (according to scalability constraints, see \autoref{discussion}). Groups (gray green full circles) are then arranged under their respective branches' ticks hanks to the \textit{Graphviz} spatialization algorithm \citep{koutsofios1996drawing}. This algorithm tries to minimize overlapping between groups and intercrossing between links. We next appoint the groups' ordinates according to their timestamps: parent groups appear at the top of the kinship view while children groups are at the bottom. Finally, we draw inheritance lines (solid dark lines) between the groups but without arrow: we think that researchers should be free to follow the natural flow of time or go back, up towards the origins of the branches. 

Still, a synchronic interpretation remains conceivable by looking horizontally at the groups. For a given year, one can observe a set of contemporary fields distributed among branches whose similarity relationships can be deduced thanks to the seabed view.

\subsection{Interacting with the views}
\label{lenses}

Since the late 1960's, the science of graphics has evolved thanks to computer sciences tools to improve data visualization. Nowadays, it is common practice to add interactive mechanisms as a way to enrich the analysis of a given graphical representation \citep{fry2004computational}. We will now describe three \textit{lenses} of exploration through which to look at the views and interact with them from macro to micro scales. 

\subsubsection{The macro lens}

The \textit{macro lens} is the default lens to explore a phylomemy as illustrated by \autoref{fig:memiescape}.1. Its first aim is to give an overview of the branches' evolution and relations of similarity. The kinship view includes zoom and drag mechanisms. By zooming in, one can focus on a given group and reveal the full name of its branch (\autoref{fig:memiescape}.5). If the user's cursor moves over a branch's tick, its corresponding groups and peak are highlighted by a yellow thread (\autoref{fig:memiescape}.1 and \autoref{fig:memiescape}.2). When a branch is dragged out of the kinship view, its peak switches off in the seabed view -- wich acts as a fixed map above the branches to prevent researchers from getting lost in the midst of the phylomemy. When one clicks on a peak, the kinship view is automatically readjusted around the coordinates of the corresponding branch.

\subsubsection{The mezzo lens}

While the \textit{macro lens} outlines the global shape of the phylomemy, \textit{mezzo~lens} focuses on its constitutive elements, namely on the emerging terms: these have been extracted beforehand (see \ref{extraction}) and are displayed in ordinate according to their date of appearance. As for the abscissa, if a term emerges in a single group we obviously reuse that group's coordinates, but if it appears twice or more at the same time, we place it at the barycenter of its emergence groups. An emerging term might also appear written in black or red according to whether it is shared by multiple branches or not. The size of the terms depends on the number of groups they are part of in the phylomemy. We also display the full list of terms one might find within a given branch when clicking on its corresponding peak (see ~\autoref{fig:memiescape}.1, right). The size of the terms appearing on the lists maps their frequency of appearance in the original corpus during the most recent period $T_{last}$. This enables us first to highlight the semantic innovations in each branch (or sets of branches)  and therefore the contributions each has made to the whole landscape; and then to show the user the vocabulary still employed at the last stage of the phylomemy. We here make the assumption that the point of view of the user is situated in time, as one usually tries to understand the current state of a given element of knowledge at~$T_{last}$ regarding its historical evolution since $T_{first}$ (e.g., while creating a bibliography). \autoref{fig:memiescape}.3 here illustrates the use of the mezzo lens to reveal major breakthroughs in the researcch conducted by the CNRS on \textit{immune response} and \textit{vaccination} between 1987 and 2004. By clicking on one of these terms, the user can switch to the \textit{micro lens}.

\subsubsection{The micro lens}

The \textit{micro lens} is designed to dive fully into the textual content of the phylomemy. It first displays terms within their respective groups (\autoref{fig:memiescape}.5) before outlining the decreasing and emerging ones with a color code: blue for decreasing terms (e.g., \textit{polymerase chain reaction} in \autoref{fig:memiescape}.6) and red for emerging terms (e.g., \textit{antibodies} in \autoref{fig:memiescape}.6). When the user clicks on a term, our phylomemy reveals the way this term among and across the branches (\autoref{fig:memiescape}.4). We put all the kinship lines linking together groups containing the targeted term in red and draw additional light red lines between any distant branches that might have been using it beforehand. In \autoref{fig:memiescape}.4, we thus highlight the shared use of the term \textit{carbon} from the branch \textit{land-cover change} (2000's - 2010's) to the branch \textit{supercapacitors} (2000's). We also bring the related branches to the front of the seabed view and add a \textit{find more} link to the Wikipedia's page of the targeted term if it exists. The micro lens thus makes it possible to follow the internal dynamics of the phylomemy and to understand trans-disciplinary influences through semantic dissemination between branches.

\subsection{Upstream extraction of a phylomemetic projection}
\label{extraction}

Views and lenses build on a set of elements that have been extracted beforehand from the high-dimensional foliation $\varphi_\delta$. We now want to detail the way we ‘slice’ into a phylomemy reconstructed in $R$ by \textit{extracting} the elements reviewed in \ref{scope}, \textit{sorting} them, \textit{filtering} them and finally \textit{labeling} them. To that end, we start by diving into the branches, so to speak, and extracting their last level of elevation with regard to the local evolution of $\delta$. The resulting network is made of branches sorted according to their drifting history (see \autoref{fig:sea_level_rise}.1). We then filter this network to remove minor branches, i.e. branches covering less than a minimal number of periods of time. This pruning aims to clarify the future reading of the visualization. Finally, we name the remaining branches by means of a two-terms label. We elect the most frequently emerging term in the targeted branch as the first component of the label. The second one is based on a classical \textit{tf–idf} measure computed within the branch's scope. If a given branch does not contain any emerging term, its label results from the union of the two terms with the highest \textit{tf–idf} score. By doing so, the branch's label should be a reasonable compromise between the specificity and the representativeness of its constitutive vocabulary. In \autoref{fig:memiescape}.2 for instance, the targeted branch is named after the union of \textit{alzheimer} and \textit{disease} and gathers research focusing on the genetical aspects of this neurodegenerative disease.

\subsection{Implementation}
\label{implementation}

In terms of technical support, the free software \textit{Gargantext} provides us with a set of fully implemented functions for the reconstruction of phylomemies. We include the most recent research developments \citep{chavalarias2020} on phylomemetic projections extraction\footnote{Code is available at \url{https://gitlab.iscpif.fr/gargantext/haskell-gargantext/tree/master}} (see \ref{extraction}) and export them as pre-spatialized Json files by means of \textit{Graphviz}. We then load those files within \textit{Memiescape}, our dedicated demonstrator for the visualization of phylomemies.

\paragraph{Gargantext.}
\label{gargantext}
It is a free text-mining software\footnote{See \url{https://gargantext.org}, \url{https://www.haskell.org/}, \url{https://reactjs.org/} \& \url{https://d3js.org/}} developed in \textit{Haskell}. Gargantext makes it possible to turn knowledge structures into tangible artifacts \cite{delanoe_mining_2020}. Gargantext addresses, by design, the user’s role in knowledge-mining tasks and therefore incorporates collaborative, cumulative and collective features. Semantic maps are created thanks to real-time peer collaboration through visualizations, easy reuse of former materials and machine learning on individual and collective past usages. The revealed shapes are consequently the outcomes of a series of reflexive choices and cumulative expertise. By using Gargantext, we aim to guarantee the easy reproducibility of our results and shorten  ‘time-to-innovation’ cycles.

\paragraph{Memiescape.} 
\label{memiescape}
It is a standalone Web demonstrator usable in a wide number of scenarios without online dependency\footnote{Code is available at \url{https://gitlab.iscpif.fr/qlobbe/memiescape/tree/v2}}. Because of scalability concerns, almost all text-mining aspects are done upstream within Gargantext. The remaining tasks are processed in the browser with JavaScript and React elements of codes to manage the views in real time. Graphics and interactive mechanisms are made in pure d3js. Memiescape is published under Gargantext licences: aGPLV3 and CECILL variant Affero compliant\footnote{See \url{https://gitlab.iscpif.fr/humanities/gargantext/blob/stable/LICENSE}}. 

%
%

\section{Results}
\label{results}

We are now able to review the reconstructed histories of thousands of scientific publications thanks to our visualization system. We rely for this on manually annotated screenshots taken from Memiescape and summarizing live explorations\footnote{We use the corpora and terms lists of \cite{chavalarias2020} available as an archive in \cite{DVN/SBH3EI_2021}}.

\subsection{An interdisciplinary corpus of academic publications}

\begin{figure}[!h]
 \centering
 \includegraphics[width=\linewidth]{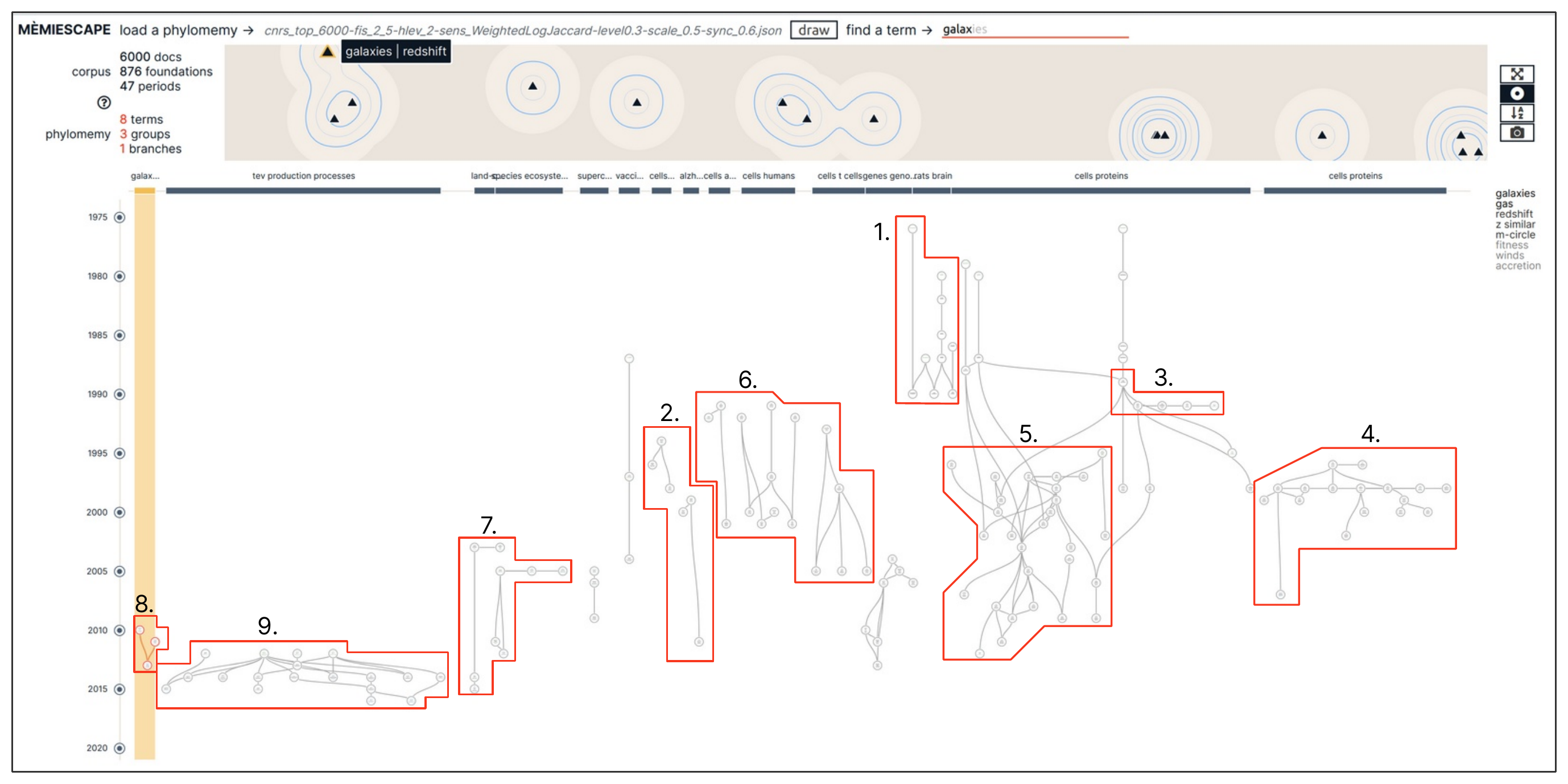}
 \caption{Manually annotated (red rectangles and arrows) phylomemy of the corpus $\mathcal{D}_{cnrs}$. Interactive version available at \url{http://maps.gargantext.org/phylo/cnrs_top_cited/memiescape/}. Phylomemy downloadable in \cite{DVN/WLI9B5_2021}.}
 \label{fig:phylo_cnrs}
\end{figure}

In section \ref{methods}, we illustrated our technical choices through the visualization of the corpus $\mathcal{D}_{cnrs}$. Interdisciplinary by nature, this corpus gathers some of the most internationally influential publications (extracted from the WoS) recently carried out by researchers affiliated to the french CNRS. Such a collection could be used as a means to understand the dynamics of research and innovation at a national scale. The collective landscape shaped by this corpus may be of interest for historians, sociologists or philosophers who investigate the underlying mechanisms of Science: academic collaboration networks, national policy effects, funding trends, etc.  Let us then go through a more detailed exploration of this corpus, using both its phylomemy (see \autoref{fig:phylo_cnrs}) and the original publications.

The reconstructed phylomemy of $\mathcal{D}_{cnrs}$ shows modern science’s global tendency to focus on the microscopic world as a mean to understand larger natural phenomenons -- from human health to biosphere changes and universe expansion. In the late 1980's for instance, neuroscience paved the way to the comprehension of \textit{brain illness} (branches no.1) by first diving into the nervous system of rats before investigating the role of the \textit{hippocampus} in \textit{memory consolidation} processes and \textit{alzheimer's disease} (branches no.2). Later on, biologists built on 1990's genomics improvements (branches no.3) to grasp how \textit{mitochondria} influenced \textit{nuclear apoptosis} mechanisms, i.e. \textit{cells death} and \textit{cancers} (branches no.4).  During the 2000's ans 2010's, genetics has led to the decoding of full \textit{genome sequences} (branches no.5) which were then used in the characterization of \textit{species} (plants, bacteria, etc.) or for treating \textit{genetic human diseases}. At the same time, medicine and pathology started to make use of genomics as well in an attempt to improve our adaptive immune system against viruses (branches no.6). Cloning techniques like \textit{monoclonal antibody} have here been applied to prevent autoimmune diseases or induce immune responses against targeted cancer cells. As for environmental researchers (branches no.7), they started to push \textit{global warming} to the fore of ecology concerns at the turn of the 2000's. They pointed out the degradation of \textit{carbon exchanges} between \textit{oceans}, \textit{tropical forests} and the whole biosphere as well as an increasing \textit{loss of biodiversity}. Beyond Earth and its atmospheric concentrations of C02, astrophysicists then tracked \textit{molecular gas} (like \textit{carbon monoxide}) and \textit{cosmic dust} (branches no.8) to discover \textit{galaxies} inside the Hubble deep field or to follow their evolution from \textit{star-forming} galaxies to mergers. Nowadays, some of the most influential CNRS publications come from the use of the \textit{large hadron collider} (branches no.9), a \textit{particle accelerator} involved in the discovery of the \textit{Higgs boson} and designed to test theoretical predictions in the fields of particle physics.

\subsection{The historical evolution of text analysis and knowledge visualization}

\begin{figure}[!h]
 \centering
 \includegraphics[width=\linewidth]{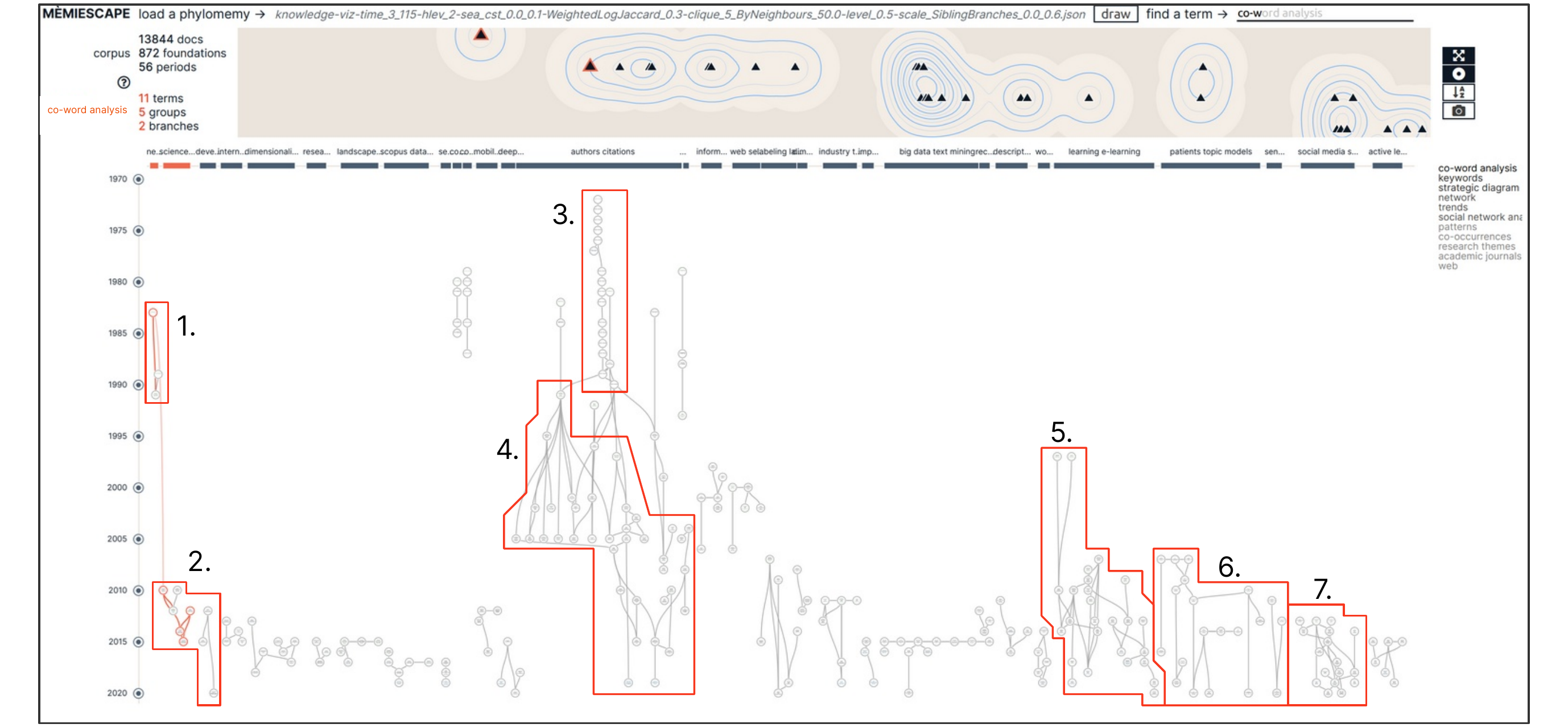}
 \caption{Manually annotated (red rectangles) phylomemy of the $\mathcal{D}_{maps}$ corpus at level 0.5, red curves highlight the spread of \textit{co-word} approaches across the branches. Interactive version available at \url{http://maps.gargantext.org/phylo/knowledge_visualization/memiescape/}.}
 \label{fig:phylo_knowledge}
\end{figure}

In section \ref{knowledge_visualization}, we used a semantic map generated by Gargantext (see \autoref{fig:map_knowledge_visualization}) to position our approach.  We will now go over the reconstruction of its corresponding phylomemy\footnote{The phylomemy $\mathcal{D}_{maps}$ can be downloaded in \cite{DVN/4FQIA9_2021}} and validate it in light of the state of the art detailed in section \ref{knowledge_visualization}. We will also add a few temporal observations.

\autoref{fig:phylo_knowledge} first outlines the evolution of \textit{co-occurrence} and \textit{co-word} analyses \citep{terzopoulos1985co,callon1991co}. These were applied in the late 1980's (branch no.1) to study paired data within a given collection of documents and, more specifically, pairs of terms for \textit{co-word} approaches. Both paradigms then enjoyed a revival of interest in the mid-2000's (branches no.2) as a result of the ICT revolution. They aimed to reveal the structural and dynamical evolution of elements of knowledge by focusing on temporal trends as well as paradigm shifts in science and research fronts \citep{chavalarias2013phylomemetic,delanoe2014dematerialization,chavalarias2020}. Our phylomemies are, in a way, heirs to these paradigms. \autoref{fig:phylo_knowledge} also shows that the classical field of \textit{citation analysis} was predominant during the 1970's (branches no.3) before passing the baton to what would become the core of \textit{bibliometry} and \textit{scientometry} in the early 1990's (branches no.4); who in turn took advantage of the emergence of large \textit{scientific databases} and new \textit{web} resources to investigate the fields of \textit{co-citation analysis} and \textit{bibliometric indices}. In the 2000's, \textit{information retrieval} techniques started to be actively used and, at the same time, the long-established field of \textit{concept mapping} found concrete applications in the domains of \textit{education} and \textit{learning process} (branches no.5). A few years later, \textit{topic modeling} rose and quickly disseminated accross various scientific fields (branches no.6), from \textit{patents} analysis to \textit{recommendations} systems and exploration of \textit{social media footprints} (branches no.5).

\subsection{A collection of timestamped clinical trials}

\begin{figure}[!h]
 \centering
 \includegraphics[width=\linewidth]{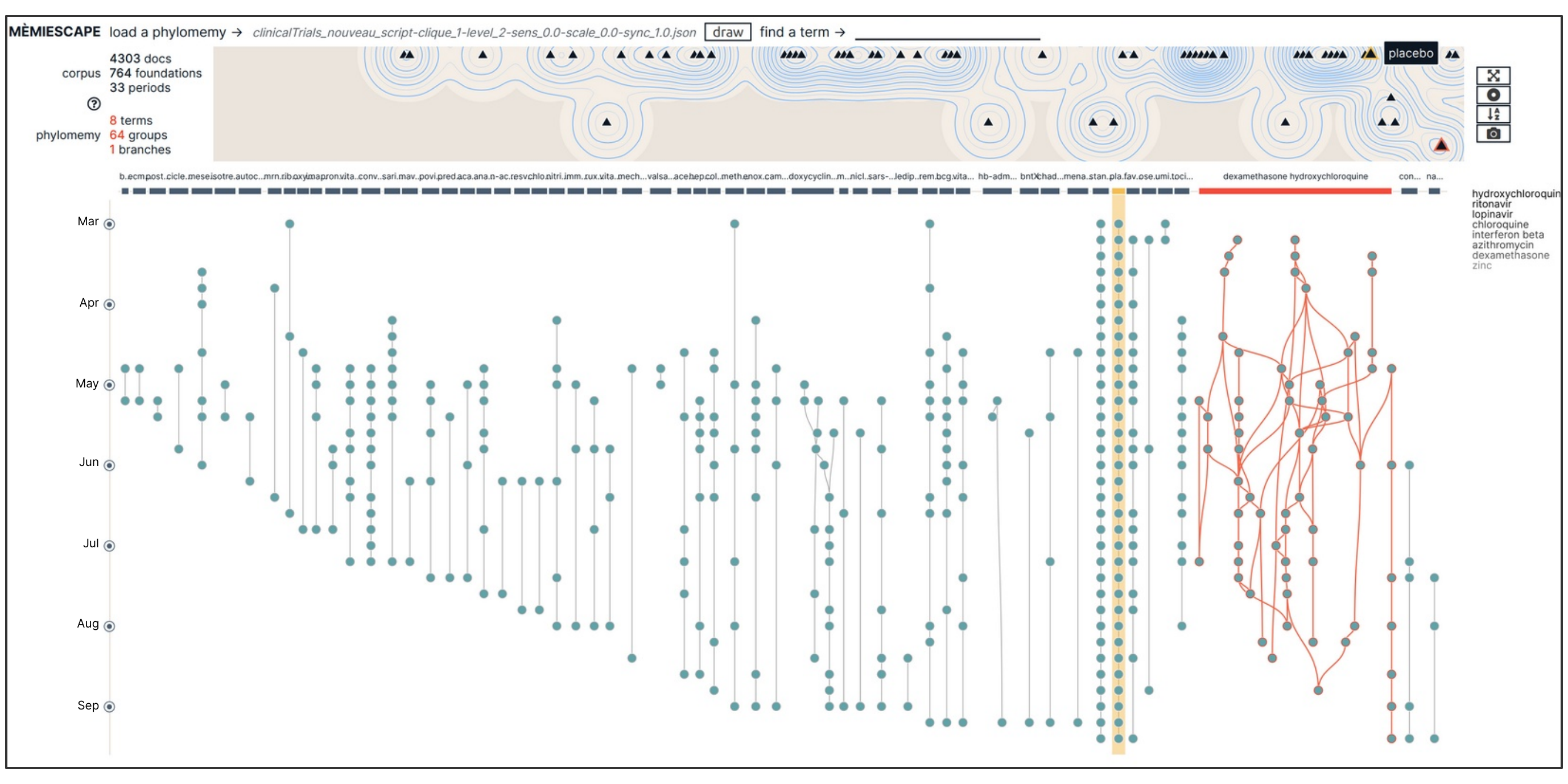}
 \caption{Phylomemy of the $\mathcal{D}_{ct}$ corpus, with \textit{hydroxychloroquine} and \textit{placebo} branches highlighted in red and yellow. Interactive version available at \url{http://maps.gargantext.org/phylo/clinical_trials/memiescape/}.}
 \label{fig:phylo_clinical_trials}
\end{figure}

As a last casework, we will now look at the reconstruction and evolution 4303 timestamped descriptions of clinical trials related to the Covid-19 pandemic. By doing so, we aim to point out that the visualization of phylomemies can apply to a wide range of both contexts and granularities of time. Here, medical descriptions have been dated according to their week of deposit within the \textit{Covid-19 WHO database}\footnote{The original database can be downloaded here \url{https://www.who.int/emergencies/diseases/novel-coronavirus-2019/global-research-on-novel-coronavirus-2019-ncov}} (from March 2020 to September 2020) and later compiled into a textual corpus named $\mathcal{D}_{ct}$\footnote{The phylomemy can be downloaded as an archive in \cite{DVN/SQULXL_2021}}. The phylomemy reconstruction summarized by \autoref{fig:phylo_clinical_trials} displays all the different research paths, discoveries and trials connected to the Covid-19 outbreak. We can see for example that the \textit{placebo} (yellow branch) has been used as a neutral element for testing the effect of different medications throughout the year 2020. More generally, \autoref{fig:phylo_clinical_trials} bears witness of a worldwide effort to find an effective cure. What is revealed here is the very making of science as an ongoing process. For instance, the branch highlighted in red translates the complex and bushy paths of \textit{chloroquine} and \textit{hydroxychloroquine} clinical trials subsequently associated to \textit{tocilizumab}, \textit{oseltamivir} or \textit{ritonavir}. It is our belief that the visualization of phylomemies could be a powerful tool to foster collective coordination between researchers.

%
%

\section{Discussion}
\label{discussion}

\subsection{A multi-scale methodology of exploration}

We have to admit that screenshots taken from \autoref{fig:phylo_cnrs}, \ref{fig:phylo_knowledge} and \ref{fig:phylo_clinical_trials} fail to translate the way we actually navigate through a phylomemy. Future improvements should therefore address the question of how to effectively translate the outcomes of an exploration in a static illustration. Yet, we think that what interactions between our views (see \ref{views}) and lenses (see~\ref{lenses}) already exist are worthwhile foundations for a convincing methodology of exploration. Users in particular have emphasized in their feedback how phylomemies' multi-scale properties stimulated their curiosity and made them want to dive deeper. Latest additions are a search engine to locate specific terms within the phylomemy and a contextual list highlighting all the terms connected to the one selected, sorted by frequency of use.

\begin{figure}[!h]
 \centering
 \includegraphics[width=\linewidth]{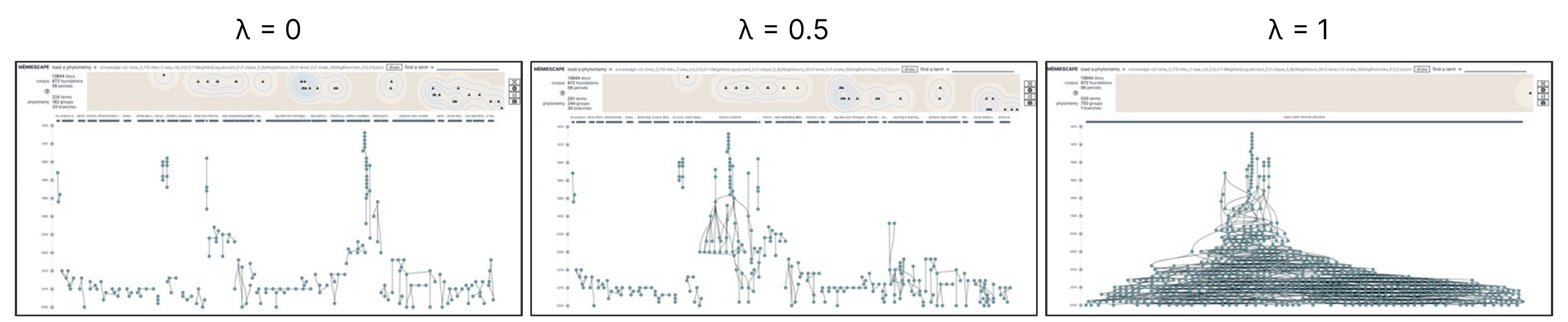}
 \caption{The phylomemy of the corpus $\mathcal{D}_{maps}$ reconstructed and visualized for three different levels of observation such as $\lambda = {0,0.5,1}$}
 \label{fig:levels}
\end{figure}

We have been inspired by the science of complex systems, in which the researcher can switch between \textit{micro}, \textit{mezzo}, and \textit{macro} scales. Here, the macro lens (see \autoref{fig:memiescape}.1) gives an overview of the temporal evolution, the mezzo lens (see \autoref{fig:memiescape}.3) helps to characterize the branches and the micro lens (see \autoref{fig:memiescape}.4) reveals the underlying semantic structure. Natural systems and phenomenons are indeed often composed of elements interacting from one scale to another. Different elements might be relevant for different scales and, for instance, micro relationships (e.g. terms used in the same document) might induce the emergence of macro shapes (e.g. branches of knowledge). Multiple scales must therefore be taken into consideration by anyone wishing to explore a phylomemy for both qualitative and quantitative analysis. Last but not least, distinct phylomemies reconstructed at different levels of observation $\lambda$ (see~\ref{phylomemy_reconstruction}) can be jointly explored in \textit{Memiescape} to reveal the whole spectrum of specialization covered by a research domain \cite{chavalarias2020}, as illustrated by the \autoref{fig:levels}. 

\subsection{Scalability issues}

Sections \ref{methods} and \ref{results} focused on the scalability of our visualization system. Two sub-issues arise: 

Firstly, the \textit{computational scalability} is first related to algorithmic issues.  Latest work in phylomemy analysis \citep{chavalarias2020} has indeed led researchers to use a reconstruction workflow that follows an $O(n^2)$ complexity. While the size of the original corpus does not seem to be a limitation, the number of terms present in the root list (see \ref{phylomemy_reconstruction}) might cause a lengthening of computation time. As for our visualization concerns, we haven't identified any complexity issue so far. But we have to notice some limitations in our current implementation: when Memiescape tries to visualize a phylomemy greater than 1000 groups, the \textit{Firefox} browser slows down and struggles to display all the graphical elements. On that particular point, we are confident that future technical developments might improve the capacity of Memiescape. 

Secondly, the \textit{graphical scalability}, related to design issues. In \autoref{fig:phylo_cnrs} or \autoref{fig:phylo_knowledge}, we've noticed that large branches' names sometime overlapped in the kinship view. But the name-shortening mechanism already implemented does not totally solve this issue. As a possible answer, we might later propose a zoom technique based on the importance of the branches or try to develop a non overlapping spatialization method for names and texts.

%
%

\section{Conclusion}

Multi-level and multi-scale by nature, a phylomemy is a complex object that winds up in a high dimensional space called $R$ \citep{chavalarias2020}. The originality of our contribution has been to propose a visualization method to endogenously project a given phylomemy from $R$ to $V$ by means of graphical views and interactive mechanisms (see \ref{methods}). The resulting system of visualization and its implementation have then been successfully applied to a wide range of elements of knowledge to browse their inner dynamics and structure (see \ref{results}). But the continuity between the original corpus of documents and its final representation can yet be improved. We should in time be able to create a flowing link connecting any term of the kinship view with its corresponding timestamped publication and, by doing so, to complete our multi-scale methodology of exploration (see \ref{discussion}).

With that in mind, we think that merging our visualization approach with the reflexive and collaborative features of a software like Gargantext \citep{delanoe_mining_2020} will make the exploration become an active process and one that allows researchers to experience the tangible nature of textual data. Since we consider phylomemies as artifacts of the researcher’s perception (see \ref{phylomemy_reconstruction}), we want to give him/her even more of a central role, set him/her in motion among the original corpora, project him/her through the whole reconstruction process. Future works will therefore be dedicated to the investigation of \textit{tangible exploratory data analysis}: a new ‘doorway’ methodology for the exploration and visualization of the hidden structure and dynamics of knowledge. This notion will question both the current and the upcoming shapes of a phylomemy thanks to a continuum of iterative loops of analysis. What is the nature of the semantic landscape I'm browsing through? What collection of documents could be missing? What is hidden beyond the borders of my corpus? What new branch of knowledge could appear if I enrich this lineage with an other? What innovative concept could emerge in a near future? Embedded within Gargantext, phylomemies will become open playgrounds where researchers are free to experience multiple round trips from the constitution of their corpora to the collaborative annotation of their visualizations.


\bibliographystyle{cas-model2-names}

\bibliography{phylomemy-viz-biblio}

%
%

\appendix

\section{The literature of text analysis and knowledge visualization}
\label{appendix_literature}

In order to extract the scientific literature of \textit{text analysis and knowledge visualization} from the \textit{Web of Science}'s online database, we have used the following queries: 

\textit{"mind map" OR "topical map" OR "knowledge map" OR "science map" OR "science mapping" OR "mapping science" OR "mapping of science" OR "semantic map" OR "co-word" OR "co-citation" OR cocitation OR "co-term" OR "concept map" OR "information cartography" OR "mapping research" OR "visualization of knowledge" OR "bibliographic coupling" OR "citation analysis" OR "topic modeling" OR "latent dirichlet"  OR (("text-mining" OR "text-analytics") AND (visualization OR infoviz OR "visual analytics"))}.

And retrieved the meta-data from a corpus of 13844 papers published between the '80s and April 2020. Please note that before 1990, most of the time, abstracts are missing in the meta-data.

\section{Data sets}
\label{appendix_data_sets}

All the corpora and lists of terms used in this paper have been described in \cite{chavalarias2020} and are available in \cite{DVN/SBH3EI_2021}. The reconstructed phylomemies can be downloaded as archives: $\mathcal{D}_{cnrs}$ in \cite{DVN/WLI9B5_2021}; $\mathcal{D}_{maps}$ in \cite{DVN/4FQIA9_2021};  $\mathcal{D}_{ct}$ in \cite{DVN/SQULXL_2021}.

\section{Acknowledgements}
\label{appendix_acknowledgements}

This research was supported by the Complex Systems Institute of Paris Île-de-France (https://iscpif.fr), the \textit{EPIQUE} project (ANR-16-CE38-0002-01), the ANR FORCCAST project and the EU FuturICT 2.0 project. We warmly thanks Bruno Gaume for his fruitful comments on our work.

\end{document}